\documentclass[12pt,a4paper]{amsart}
\usepackage[a4paper,margin=3cm]{geometry}

\usepackage{graphicx}
\usepackage{amsmath}
\usepackage{amssymb}
\usepackage{xurl}
\usepackage{hyperref}
\usepackage{latexsym}
\usepackage{tabularx} 
\usepackage{rotating}
\usepackage{multirow}
\usepackage{environ}
\usepackage{caption}
\usepackage{pdflscape}
\usepackage[shortlabels]{enumitem}
\usepackage{subcaption}
\usepackage{comment}
\usepackage{mathtools}
\usepackage{pgf,tikz}
\usepackage{threeparttable}
\usepackage{diagbox}
\usepackage[authoryear, round]{natbib}
\usepackage{arydshln}
\usepackage[table]{xcolor}
\usepackage{booktabs}
\usepackage{siunitx}

\newcounter{examplecounter}
\makeatletter
\newcommand{\addresseshere}{%
	\enddoc@text\let\enddoc@text\relax
}
\makeatother

\begin{document}

\title[Operationalizing Allocation Probability Tests]{Operationalizing Allocation Probability Tests: Practical Guidance on Optimized Implementation for Power and Robustness}
\author{Stina Zetterstrom\textsuperscript{1}, David S. Robertson\textsuperscript{1}, Thomas Jaki\textsuperscript{1,2}, Sof\'ia S. Villar\textsuperscript{1}}
\address{\textsuperscript{1} MRC Biostatistics Unit, University of Cambridge, United Kingdom\\
\textsuperscript{2} Faculty of Informatics and Data Science, University of Regensburg, Germany}
	\email{stina.zetterstrom@mrc-bsu.cam.ac.uk}
\date{\today}
\begin{abstract}

Recently, a new testing approach for response-adaptive clinical trials was proposed based on the allocation probabilities (AP) rather than the outcome data. While original work on the AP test focused on binary and normal endpoints and demonstrated that significant efficiency gains are possible, many critical questions remain open regarding its practical implementation and upper limits. In this work, rather than simply proposing novel statistics, we seek to understand the maximum gain that can be obtained with the AP test by optimizing how these probabilities are used to define the test statistic. We expand the method’s practical utility by applying it to survival endpoints (exponential distributions) and introducing a rigorous strategy for selecting the null hypothesis to properly calibrate type I error. Our simulation studies reveal that by optimizing the functional form of the AP test, investigators can achieve a substantial increase in power, approaching the theoretical maximum, without sacrificing the patient outcome goals of the design. Furthermore, we explicitly compare the method to a standard Bayesian decision rule, finding that the optimized AP test significantly outperforms traditional frequentist tests while maintaining strict error control. This work provides a missing practical framework for implementing robust and optimized AP tests in complex response-adaptive settings.
	 
\end{abstract}

\maketitle


\newpage
\section{Introduction}

Response-adaptive clinical trials update treatment allocation probabilities sequentially based on accumulating data, primarily to maximize patient benefit within the trial. While this may offer advantages in terms of expected patient outcomes, the resulting treatment imbalance often leads to a substantial loss in statistical power when using traditional frequentist tests. This paper focuses on the Allocation Probability (AP) test, a recent innovation that bypasses outcome-based inference by using the allocation probabilities themselves as the test statistic \citep{barnett2023novel,deliu2025efficient}. Although initial research demonstrated that the AP test can recover efficiency in binary and normal endpoint settings, many critical questions remain regarding its operational limits and practical implementation.

In this work, we seek to understand the maximum gain obtainable with the AP test by optimizing the functional forms used to define the test statistic. We move beyond simple comparisons to provide a practical framework for its use in practice, expanding the method’s utility to survival endpoints and providing a rigorous strategy for type I error calibration. By optimizing these functional forms, we demonstrate that investigators can achieve power gains that approach the theoretical maximum without sacrificing the trial’s objectives.

In more detail, we operationalize the AP test, expanding the framework across several practical dimensions to ensure robust implementation. First, we look at different types of endpoints (exponential, binary, and normal endpoints), with our main focus on the performance of the AP test for time-to-event (exponential) outcomes, which has previously not been studied. Second, we look at alternative functional forms of the AP test to see if the performance of the AP test, in terms of statistical power, can be improved by changing the weighting of the sequentially calculated allocation probabilities in the test. We demonstrate relevant power gains for alternative functional forms to the original form, and while all the ones considered did better than the original one, we find that the most powerful version of the AP test only includes the last allocation probability. Third, we find that a Bayesian decision rule is a special case of the AP test, and this, and other versions of the AP test, are compared to other hypothesis tests. We look at the performance of these versions of the AP test in comparison to other methods as different settings are varied, such as the trial and block size, the number of blocks to include in the test statistic, the parametric space, etc. We limit our investigations to two-arm trials in two different BRAR designs. The reasons we investigate the two-arm case is that it is the worst case for power in RAR trials, it is common for confirmatory settings, and it can be applied to pairwise comparisons in platform trials. We only consider different BRAR designs as BRAR is the most commonly used RAR in practice \citep{pin2025github}. In summary, our investigations are aimed at answering the questions:
\begin{enumerate}
    \item How much can the functional form of the AP test increase power for the different outcome measures?
    \item How does the most powerful version of the AP test compare to a
    \begin{enumerate}
        \item non-adaptive design with a traditional analysis;
        \item BRAR design with a frequentist analysis;
        \item BRAR design with a Bayesian analysis?
    \end{enumerate}
    \item Based on the above, can we make some practical recommendations?
\end{enumerate}
The rest of the article is outlined as follows. Section \ref{sec:prob} describes the problem setting, the BRAR algorithm, and the AP test. Section \ref{sec:func} presents a generalized functional form of the AP test and compares its performance to alternative methods in simulations and Section \ref{sec:emp} illustrates the use of the modified AP tests in an empirical example. Section~\ref{sec:guide} gives practical guidelines for how to implement to AP test, and finally, the results are discussed in Section \ref{sec:disc}.

\section{Problem setting} \label{sec:prob}

We consider a two-armed trial where some patients are allocated in a burn-in block with equal randomization (ER) of size $B'$ and the other patients are allocated to treatments in blocks of size $B$, i.e. first $B'$ are allocated as in an ER trial and the following patients are randomized according to the RAR procedure. The block sizes can change as the trial progress, but throughout this article, we assume that \textit{B} is the same for all blocks. The treatments are allocated at times $t=0,\dots,T$, with a total sample size of $N=B' + B\cdot T$. At each time/block $t$, each subject $i=1,\dots,B$ in the block is assigned to one treatment arm $A_{t,i}$, where $A_{t,i}=0$ denotes that subject \textit{i} in block \textit{t} is assigned to the control arm and $A_{t,i}=1$ denotes that subject \textit{i} in block \textit{t} is assigned to the experimental treatment arm. The treatment arms are allocated based on allocation probabilities, $\pi_{a}=\{\pi_{t,a}, a=0,1\}$, where $\pi_{t,a}$ is the allocation probability to arm \textit{a} for all subjects in block \textit{t}, such that $0<\pi_{t,a}<1$. In the two arm setting, $\pi_{t,0}=1-\pi_{t,1}$. Note that all subjects in the same block have the same probability to be assigned to the different arms. We use the Neyman-Rubin causal model \citep{Neyman23,rubin1974estimating,rubin1977assignment} to define two potential outcomes for each subject, $Y_{t,i}(A_{t,i})$, for $A_{t,i}=0,1$. The potential outcomes are the outcome that would be observed if subject \textit{i} in block \textit{t} was assigned to arm $A_{t,i}$. We make the Stable Unit Treatment Value Assumption (SUTVA), i.e., that the observed outcome is the potential outcome under the received treatment. We also assume that the potential outcomes are independent and identically distributed, meaning that the potential outcomes do not depend on $\mathcal{H}$ (defined below). The history of selected arms and outcomes up to block \textit{t} is $\mathcal{H}_{t-1}=\{A_{\tau,i},Y_{\tau,i}(A_{\tau,i}),i=1,\dots,B,\tau=0,\dots,t-1\}$, and the allocation probabilities are functions of $\mathcal{H}$: $\pi_{t,a}=\mathbb{P}(A_{t,i}=k|\mathcal{H}_{t-1})$, following the procedure detailed next. 

\subsection{Bayesian response-adaptive randomization} \label{sec:brar}

The BRAR algorithm is the earliest algorithm for response-adaptive randomization \citep{thompson1933likelihood}. It is defined in a Bayesian setting, and the allocation probabilities to each treatment arm are proportional to the posterior probabilities that the respective arms correspond to the best treatment given data. In other words, the probability to allocate a subject to the experimental treatment arm at block \textit{t} is
\begin{equation} \label{eq:TS}
    \begin{split}
        \pi_{t,1}^{BRAR}&=\mathbb{P}\left[ \mathbb{E}\{Y_{t,i}(1)\}\geq \mathbb{E}\{Y_{t,i}(0)\} \mid \mathcal{H}_{t-1}\right],
    \end{split}
\end{equation}
for $t=1,\dots,T$. The allocation probabilities are used to allocate the subjects in block \textit{t}, and are updated before the next block. For $t=T$ the allocation probabilities can be used for decision making but no patients are randomized after that. In some settings, the BRAR allocation probabilities can be calculated analytically, but often they must be obtained through numerical simulations \citep{kaddaj2025thompson}. Here, the allocation probabilities are calculated analytically. More details on the BRAR algorithm can, for example, be found in \citet{russo2018tutorial}.

When the experimental treatment is superior, the distribution of the allocation probabilities are skewed towards 1,
\begin{equation} \label{eq:conv}
    \pi_{t,1}^{BRAR} \rightarrow 1 \;\; \mathrm{as} \;\; t \rightarrow \infty
\end{equation}
\citep{kalkanli2020asymptotic}, and as a result, a large imbalance between the sample sizes of the treatment arms typically arise in finite samples. When treatment arms are identical, the distribution of $\pi_{t,1}^{BRAR}$ does not converge \citep{zhang2020advances}, but imbalanced sample sizes may arise anyway. 

The standard BRAR design can be very reactive in favoring an arm early in the trial and thus prematurely favor an incorrect arm. One way to avoid this is to regularize the allocation probabilities, i.e., instead of randomizing according to the posterior probabilities in \eqref{eq:TS}, use probabilities
\begin{equation} \label{eq:tunedTS}
    \pi_{t,1}^{BRARc}=\frac{\left(\pi_{t,1}^{BRAR}\right)^c}{\left(\pi_{t,1}^{BRAR}\right)^c+\left(\pi_{t,0}.^{BRAR}\right)^c}
\end{equation}
This type of design would still, on average, favor the better arm (when it exists), however less fast than the original BRAR design. In this work, we will use the regularized version of BRAR with $c=0.1+0.9\cdot t/T$ so that the algorithm is more explorative early in the trial and more exploitative later in the trial, but it can never be more aggressive than the untuned BRAR algorithm. 

Both the standard BRAR design and the tuned BRAR can result in a large imbalance in sample size between the treatment arms. This is problematic for traditional methods for estimation and hypothesis testing, and alternative approaches is needed. One such alternative hypothesis test is the AP test.

\subsection{Allocation probability test} \label{sec:ap}

The AP test was suggested in \citet{barnett2023novel} for binary outcome with an FLGI design. \citet{deliu2025efficient} also studied the AP test for normal outcome with equal variance with a BRAR design. In the two-arm setting, the AP test statistic is defined as
\begin{equation} \label{eq:AP}
    \mathrm{AP}_T=\sum_{t=t_{\mathrm{min}}}^{T+1} \mathbb{I}(\pi_{t,1}>0.5),
\end{equation}
where $t_{\mathrm{min}}$ is the first block from when the adapted probabilities are included in the test statistic. The minimum is $t_{\mathrm{min}}=1$, i.e. the first block after the burn-in period $B'$ with updated allocation probabilities, since the allocation probabilities in the test must be based on some previous data. It is however possible to start adapting the allocation probabilities but not include them in the test statistic, i.e. use a run-in period, and have $t_{\mathrm{min}}>1$. Extension of the AP test to multi-arm trials (with \textit{M} experimental arms) have also been presented, and the main difference to the two-arm case is that 0.5 in the AP test statistic is replaced with $1/(M+1)$. We introduce a subtle but important departure from previous definitions, which restricted their scope to observed blocks up to $T$. Here, we extend the definition to include the allocation probabilities for the hypothetical final block at time $T+1$. Even though patients are not physically allocated in this subsequent block, the allocation probabilities calculated for it capture significant residual information about the treatment effect.

The idea behind the test is that the allocation probabilities will favor the most promising treatment arm, especially in a more reactive response-adaptive design. The allocation probabilities will therefore differ from the allocation probabilities of an ER design, which are 0.5 in a two-arm case. Using the result in \eqref{eq:conv}, \citet{deliu2025efficient} shows that when $T\rightarrow \infty$
\begin{equation}
    \mathrm{AP}_T\rightarrow\infty
\end{equation}
when the allocation probabilities are defined according to a standard BRAR and the arms are non-identical.

To test a null versus alternative hypotheses, the distribution of the AP test statistic under the null and an associated critical value,
\begin{equation}
    q_{\alpha}=\min\{q\in[0,\dots,T]:\mathbb{P}(\mathrm{AP}_T>q|H_0)\leq \alpha \}
\end{equation}
must be derived. In some cases these probabilities can be calculated exactly, but often simulations must be used. Note that the distribution and critical value depend on the number of blocks, \textit{T}. Furthermore, the AP test is discrete and it may therefore be impossible to use standard significance levels such as $\alpha=0.05$, without using randomized tests. For most realistic trial sizes ($\lesssim 1000$), achieving a standard significance level requires using the maximum possible critical value, $q_\alpha = T + 1 - t_{min}$. Any threshold strictly greater than this results in a test that can never reject the null hypothesis~\citep{deliu2025efficient}.

Earlier works have shown that the original form of the AP test in \eqref{eq:AP} can achieve power gains compared to traditional methods \citep{barnett2023novel,deliu2025efficient}. Here, we investigate the performance of the AP test in more detail with a focus on the power improvements of the approach via the functional form of the test statistic.

\section{Analysis of the generalized AP test} \label{sec:func}

\subsection{The generalized AP test} \label{sec:apfunc}

The AP test in \eqref{eq:AP} weights all blocks with $\pi_{t,1}>0.5$ equally regardless of the size of $\pi_{t,1}$ or the value of $t$. However, different blocks may contain different amount of information, perhaps resulting in more power gains if the different blocks are not weighted equally. In other words, the RAR information in the trajectory can be reflected by weighting the blocks. Furthermore, this information can also be incorporated by not dichotomizing the allocation probabilities. There are several ways to weight the blocks and  several alternatives to the indicator function. To take these things into account, we introduce the generalized AP test statistic, defined as
\begin{equation} \label{eq:APgen}
    \mathrm{AP}_T^{f,w}=\sum_{t=t_{\mathrm{min}}}^{T+1} f(\pi_{t,1})\cdot w_t.
\end{equation}
The original AP test had $f(\pi_{t,1})=\mathbb{I}(\pi_{t,1}>0.5)$ and $w_t=1$. We propose and investigate some versions of $f(\cdot)$ and $w_t$, although several others can be justified. 

One way to weight the blocks is based on time, $w_t=t$, $t=1,\dots ,T+1$, with the justification that the allocation probabilities for blocks later in the trial are based on more data points and may therefore contain more information. Furthermore, the AP test can be based on the allocation probabilities directly, without the indicator function, $f(\pi_{t,1})=\pi_{t,1}$. This approach also takes the size of the allocation probabilities into account. By combining these two alternatives of $w_t$ and $f(\pi_{t,1})$, we define the timedirect AP test,
\begin{equation}
    \mathrm{AP}_T^{timedirect}=\sum_{t=t_{\mathrm{min}}}^{T+1} \pi_{t,1}\cdot t.
\end{equation}
We also consider the alternative AP test with $f(\pi_{t,1})=\pi_{t,1}$,$w_t=0$ for $t=1,\dots T$, and $w_{T+1}=1$, and we define the last-block AP test,
\begin{equation}
    \mathrm{AP}_T^{lastblock}= \pi_{T+1,1}.
\end{equation}
When a BRAR design is used, this version of the AP test naturally reduces to a Bayesian decision rule.

These are two of the possible different versions of the AP test. Here, we only present the timedirect and last-block AP tests, and the definitions of the other alternatives are found in the Supplemental Material.

\subsection{Simulation studies} \label{sec:sim}

We follow the ADEMP framework from \cite{morris2019using} in the presentation of the simulation design.  ADEMP stands for \textit{Aims}, \textit{Data-generating mechanisms}, \textit{Estimands}, \textit{Methods}, and \textit{Performance measures}. Here, hypothesis tests instead of an estimand are of interest, and following the authors' notation, we use the term \textit{Targets} instead of \textit{Estimands}.

\textbf{Aims:} To (1) evaluate the overall performance of the new versus old functional form for the AP test proposed in Section~\ref{sec:apfunc} for two different versions of BRAR and (2) compare both version of the AP test to other methods and designs and (3) evaluate in which settings (a version of) the AP test is appropriate to use.

\textbf{Data-generating mechanisms:} We evaluate the methods in settings that mimic phase 2 and phase 3 studies. For the phase 2 simulations, the sample size is $N=100$, and for the phase 3 simulations, $N=500$. These sample sizes are chosen to achieve a power in an ER design higher than 60\% and 90\% when the treatment effect is 50\%. The phase 2 study has $B'=10$ and $B=1$, and the phase 3 study has $B'=50$ and $B=10$. Different to past works, we consider a time-to-event setting such that a higher rate (which equals a shorter time to event) is beneficial, i.e. time to recovery or something similar. The outcome data follows an exponential distribution without censoring, $Y\sim Exp(\lambda_a)$, where $\lambda_0=1$ and $\lambda_1=1.2,\;1.4,\;1.6,\;1.8,\;2.0$. The designs that are considered are the standard BRAR design and the tuned BRAR design, presented in Section~\ref{sec:brar}. Both designs are also compared against a baseline ER design which is a permuted block design with block size 8, with the last block being the leftover to get to the total sample size. 

\textbf{Targets:} We consider the hypotheses:
\begin{equation} \label{eq:hyp}
    H_0: \;\lambda_1=\lambda_0=1 \;\; \mathrm{vs} \;\; H_1: \;\lambda_1 >\lambda_0 \;\; \mathrm{with} \;\; \lambda_1=1.2,\;1.4,\;1.6,\;1.8, \;2.0.
\end{equation}

\textbf{Methods:} Each simulated dataset from one of the BRAR designs is analyzed using the following hypothesis tests:
\begin{enumerate}
    \item Likelihood-ratio test
    \item Original AP test; $w_t=1$ for $t=1,\dots,T+1$ and $f(\pi_{t,1})=\mathbb{I}(\pi_{t,1}>0.5)$.
    \item Timedirect AP test; $w_t=t$ for $t=1,\dots,T+1$ and $f(\pi_{t,1})=\pi_{t,1}.$
    \item Last-block AP test; $w_t=0$ for $t=1,\dots,T$, $w_{T+1}=1$, and $f(\pi_{t,1})=\pi_{t,1}.$ In these simulation studies, this version of the AP test is identical to a Bayesian decision rule using the  posterior probabilities $\pi_{T+1,1}^{BRAR}=\mathbb{P}(\lambda_1>\lambda_0\mid \mathcal{H}_{T})$ with a cut-off value such that 5\% type I error rate is achieved when the arms are identical and equal to the true null.
\end{enumerate}
These are also compared to a likelihood-ratio test from the permuted block ER design. Results for additional alternative versions of the AP test are presented in the Supplemental Material.

\textbf{Performance measures:} We compare type I error rate, power, proportion of patients on the better treatment, and mean time to outcome. Our key performance measures are type I error rate and power. We also report Monte Carlo standard errors, which we require to be $\leq 0.005$. We use $10^6$ replicates for the calibration of the critical values (to achieve the targeted type I error rates under $H_0$) and $10^5$ replicates for the other simulations.

For both treatment arms in both designs, we use a $Gamma(1,0.001)$ prior distribution, which is vague prior. The posterior distribution for treatment arm $k$ is $Gamma(1+n_k,0.001+\sum y_k)$, where $n_k$ is the total number of patients allocated to arm $k$ and $\sum y_k$ is the total time until event on that treatment arm. Using an integer value for the first parameters makes it possible to calculate the posterior probabilities in \eqref{eq:TS} exactly which decreases the runtime significantly. Note that the critical value of the AP tests and the Bayesian decision rule for the null hypothesis in \eqref{eq:hyp} are found using simulations. With the prior distribution and outcome model, the chosen BRAR design does not depend on the exact values of the rate parameters, only the ratio between them, i.e., there is no difference in the BRAR design if $\lambda_0=\lambda_1=1$ or if $\lambda_0=\lambda_1=10$. Thus, the critical values for the AP tests and Bayesian decision rule will not differ regardless of which parameter values they are calibrated under. Note that this is an exception and does not hold for other types of outcome.

\subsubsection{Phase 2}

In a phase 2 trial, the sample size is typically relative small, and it is desired to maximize the power as long as the type I error rate is kept to a reasonable level, usually 10\% \citep{rubenstein2005design}. Therefore, we look at power with a targeted 10\% significance level, although we do not adjust the hypothesis tests to keep it fixed at that level, i.e. the type I error rate is close to 10\%.

\begin{figure}[htb!]
\centering
\begin{subfigure}[t]{0.49\linewidth}
    \centering
    \includegraphics[width=\linewidth]{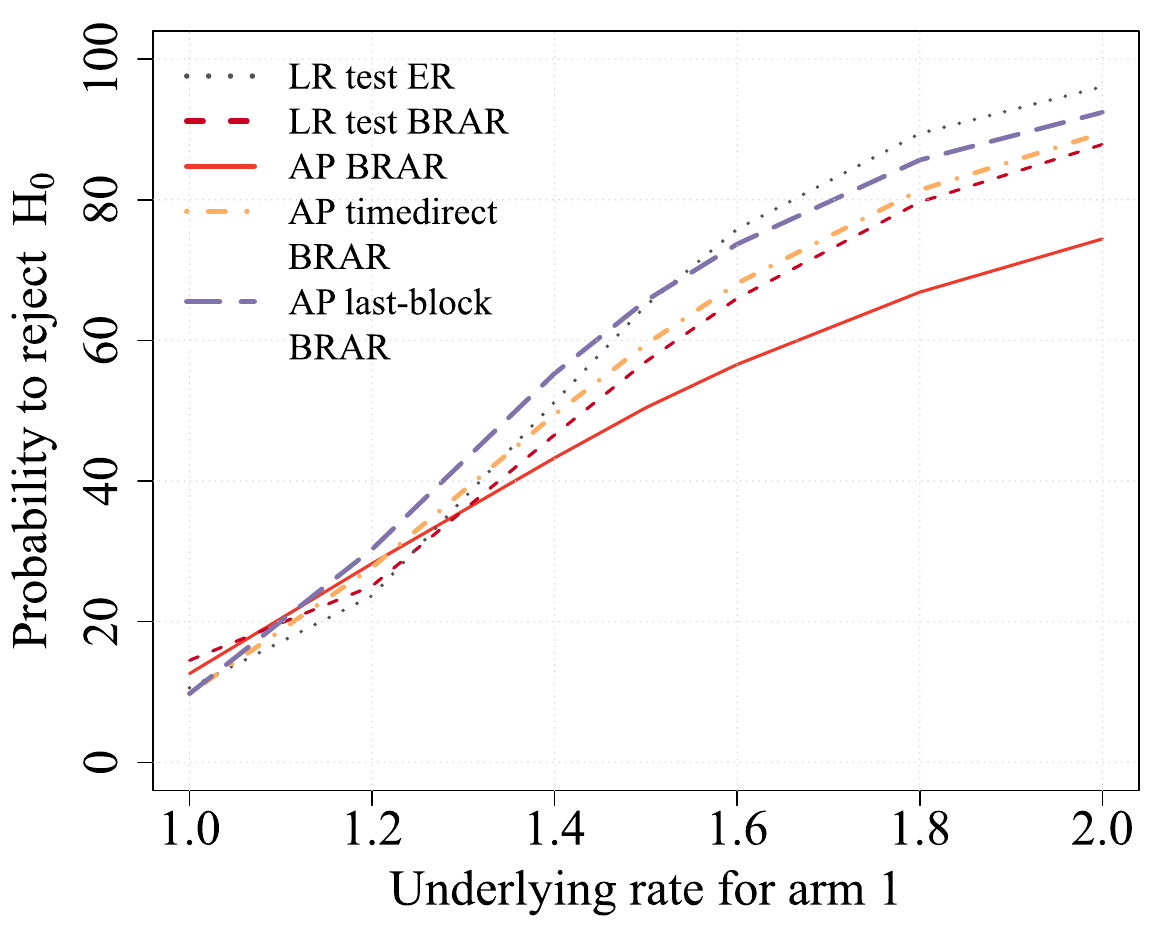}
	\caption{Standard BRAR.}
    \label{fig:standardTS2}
\end{subfigure}
\begin{subfigure}[t]{0.49\linewidth}
    \centering
    \includegraphics[width=\linewidth]{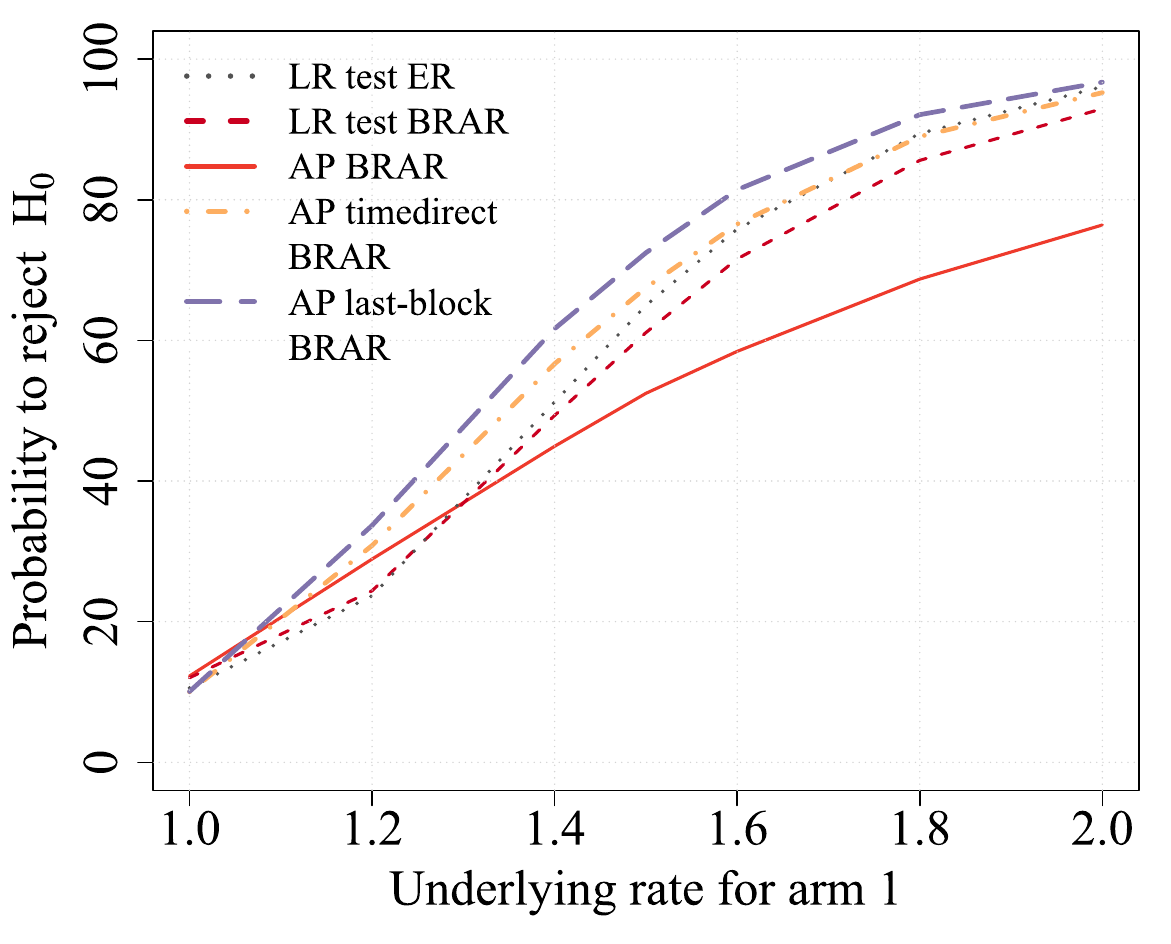}
	\caption{Tuned BRAR.}
    \label{fig:tunedTS2}
\end{subfigure}
\caption{Power curves for (A) standard BRAR and (B) tuned BRAR for $N=100$ and $B=1$, Monte Carlo standard errors $\leq 0.005$.}
\label{fig:powerPhase2}
\end{figure}

In Figure~\ref{fig:powerPhase2}, we see the power curves for the different tests. For both designs, we see that changing the functional form of the AP test statistic can increase the power of the test, without changing anything in the design. Specifically, the last-block AP test has the highest power of the tests for the BRAR designs, but we also see that both the last-block and timedirect AP tests have higher power than the likelihood-ratio test for the two different BRAR designs. Comparing the two designs, we see that both the last-block  and timedirect AP tests have higher power for the tuned BRAR design. Furthermore, the last-block and timedirect AP tests for the standard BRAR design have either slightly lower or higher power than the likelihood-ratio test for the ER design, depending on the parameter space, whereas for the tuned BRAR design, they have higher or equal power. It should also be noted that the last-block and timedirect AP tests have the targeted type I error rate of 10\% whereas the type I error rate of the likelihood-ratio test for the BRAR designs is larger than 10\%. The original AP test has the highest power for the lower parameter values, although that is largely due to a type I error rate of around 13\%, which is larger than the targeted 10\%.  

The expected patient benefit is compared for the BRAR and ER designs in Table~\ref{tab:patben}. When the treatment ratio is 50\% it can be seen that the average percentage of patients on correct treatment (i.e. the experimental arm) increases from 50\% in an ER design to 73\% for the tuned BRAR design and 79\% for the standard BRAR design, and that the average time to outcome decreases from 0.81 in the ER design to 0.76 for the tuned BRAR design to 0.74 for the standard BRAR design.

\subsubsection{Phase 3}

In a phase 3 trial, the sample size is typically larger, and the significance level is set by the regulators, usually at 5\%. This represents an absolute constraint; statistical power cannot be gained by compromising the type I error rate. Therefore, we look at power when the hypothesis tests are adjusted to strictly control the type I error at a 5\% significance level.

\begin{figure}[htb!]
\centering
\begin{subfigure}[t]{0.49\linewidth}
    \centering
    \includegraphics[width=\linewidth]{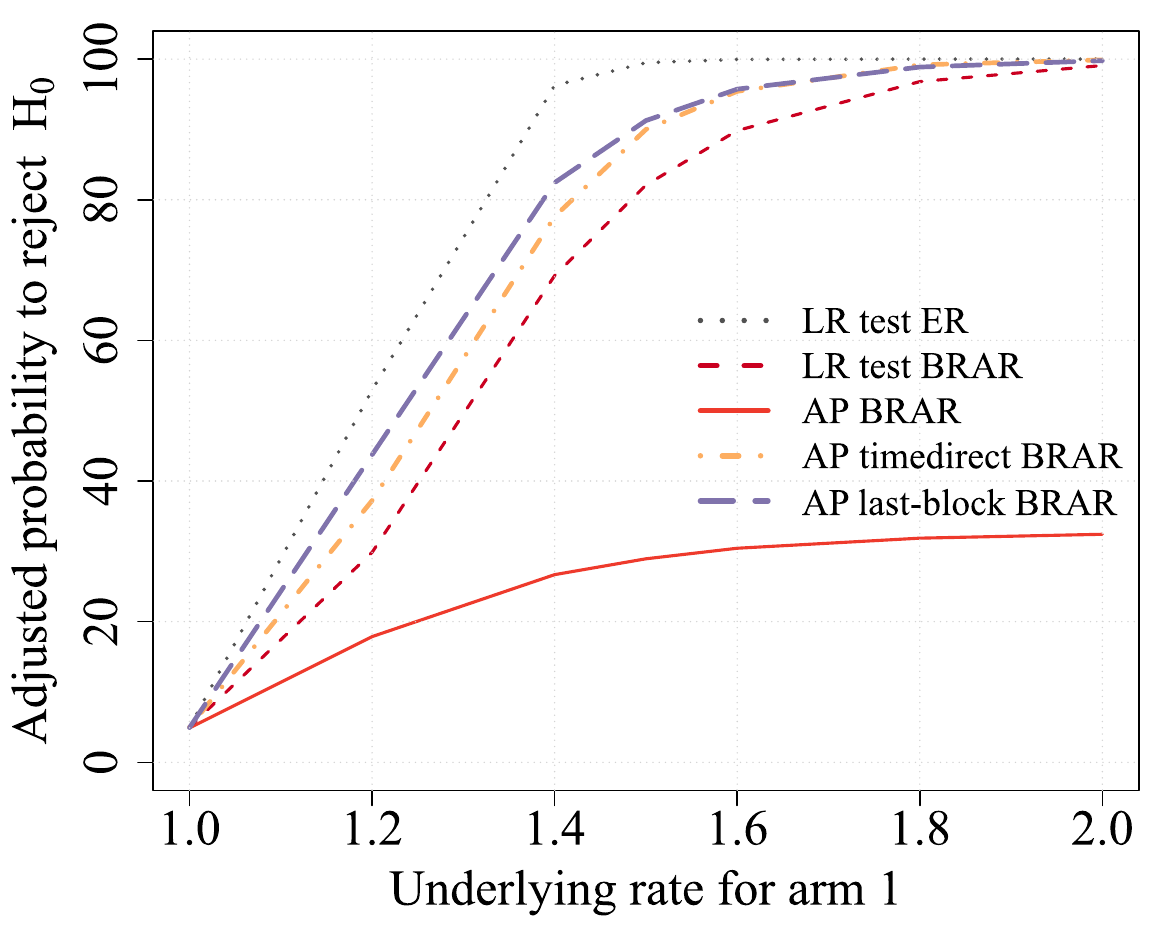}
	\caption{Standard BRAR.}
    \label{fig:standardTS3}
\end{subfigure}
\begin{subfigure}[t]{0.49\linewidth}
    \centering
    \includegraphics[width=\linewidth]{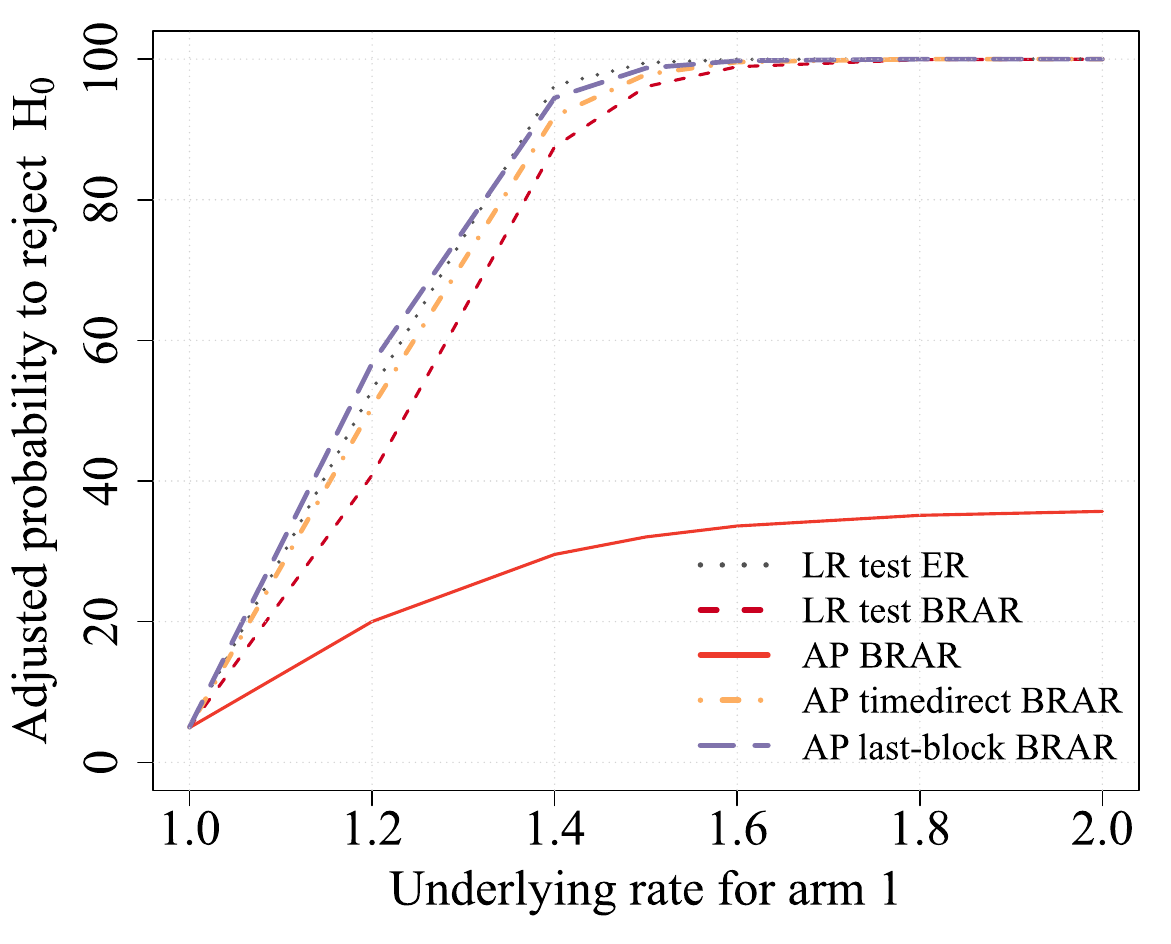}
	\caption{Tuned BRAR.}
    \label{fig:tunedTS3}
\end{subfigure}
\caption{Power curves for (A) standard BRAR and (B) tuned BRAR for $N=500$ and $B=10$ with adjustment for strict type I error control, Monte Carlo standard errors $\leq 0.005$.}
\label{fig:adjpowerPhase3}
\end{figure}

For both designs in Figure~\ref{fig:adjpowerPhase3}, we see that the last-block AP test has the highest power of the tests for the BRAR designs. Furthermore, we see that the last-block and timedirect AP tests have large power gains, between 20\% and 50\%, compared to the original AP test. The original AP test has a very low power, even for large values of $\lambda_1$. This comes from the high type I error rates without the adjustment to strict type I error control. Since the type I error rate without adjustment is roughly three times as large as the desired 5\% level, even a 100\% power without adjustment will result in around 30\% power with adjustment. The last-block and timedirect AP tests have power gains compared to the likelihood-ratio test for both BRAR designs for all parameter values. Similar to the phase 2 simulations, the last-block and timedirect AP tests have higher power in the tuned BRAR design than in the standard BRAR design. For the standard BRAR design, they both have lower power than the likelihood-ratio test for the ER design, but for the tuned BRAR design they have similar power to the likelihood-ratio test for the ER design. Furthermore, the gain in patient benefit, seen in Table~\ref{tab:patben}, must also be taken into account. Here, we see that the percentage of patients on the correct treatment increases from 50\% in the ER design to 86\% for the tuned BRAR design to 91\% for the standard BRAR design and the average time to outcome decreases from 0.81 for the ER design to 0.72 for the tuned BRAR design to 0.70 for the standard BRAR design in the phase 3 simulations. 

\subsubsection{Type I error rates}

\begin{figure}[htb!]
\centering
\begin{subfigure}[t]{0.49\linewidth}
    \centering
    \includegraphics[width=\linewidth]{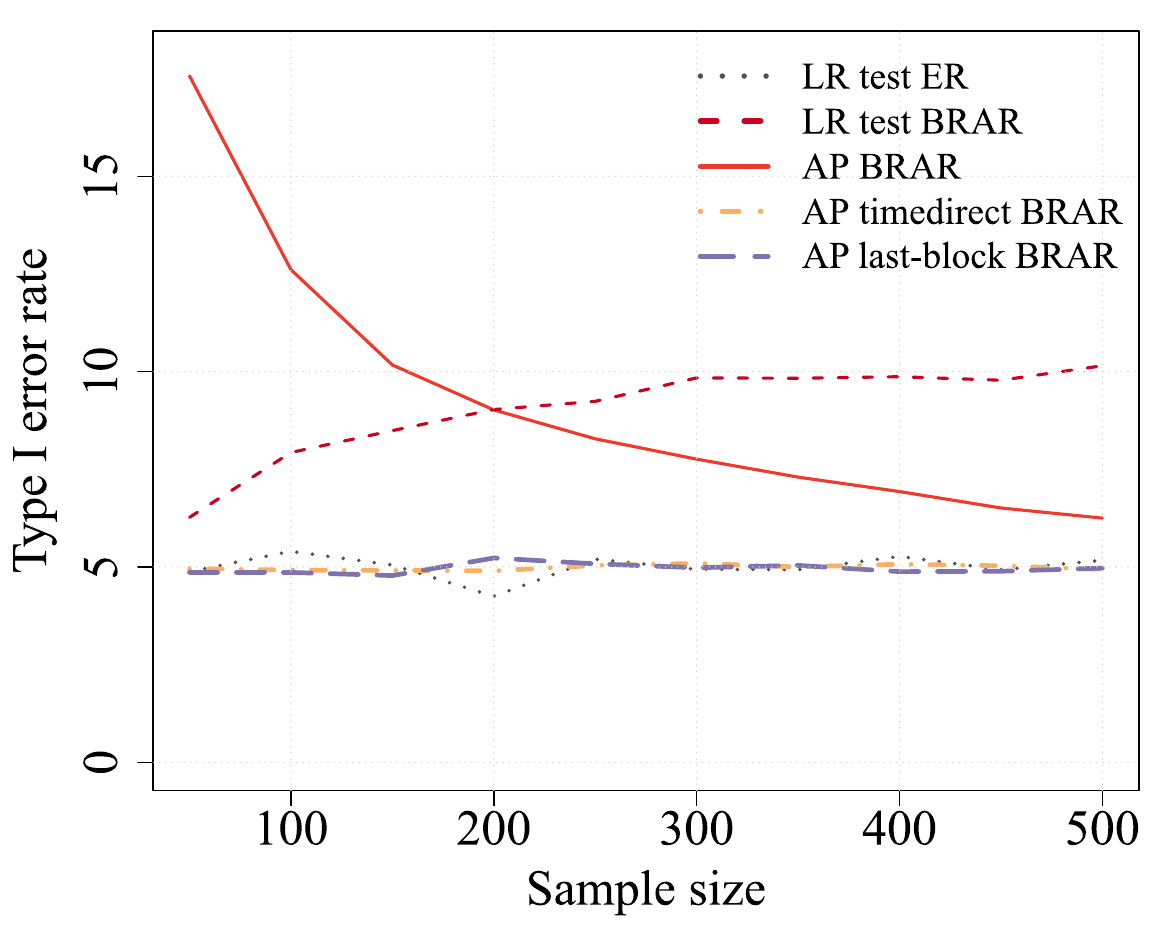}
	\caption{Standard BRAR.}
    \label{fig:standardTST1}
\end{subfigure}
\begin{subfigure}[t]{0.49\linewidth}
    \centering
    \includegraphics[width=\linewidth]{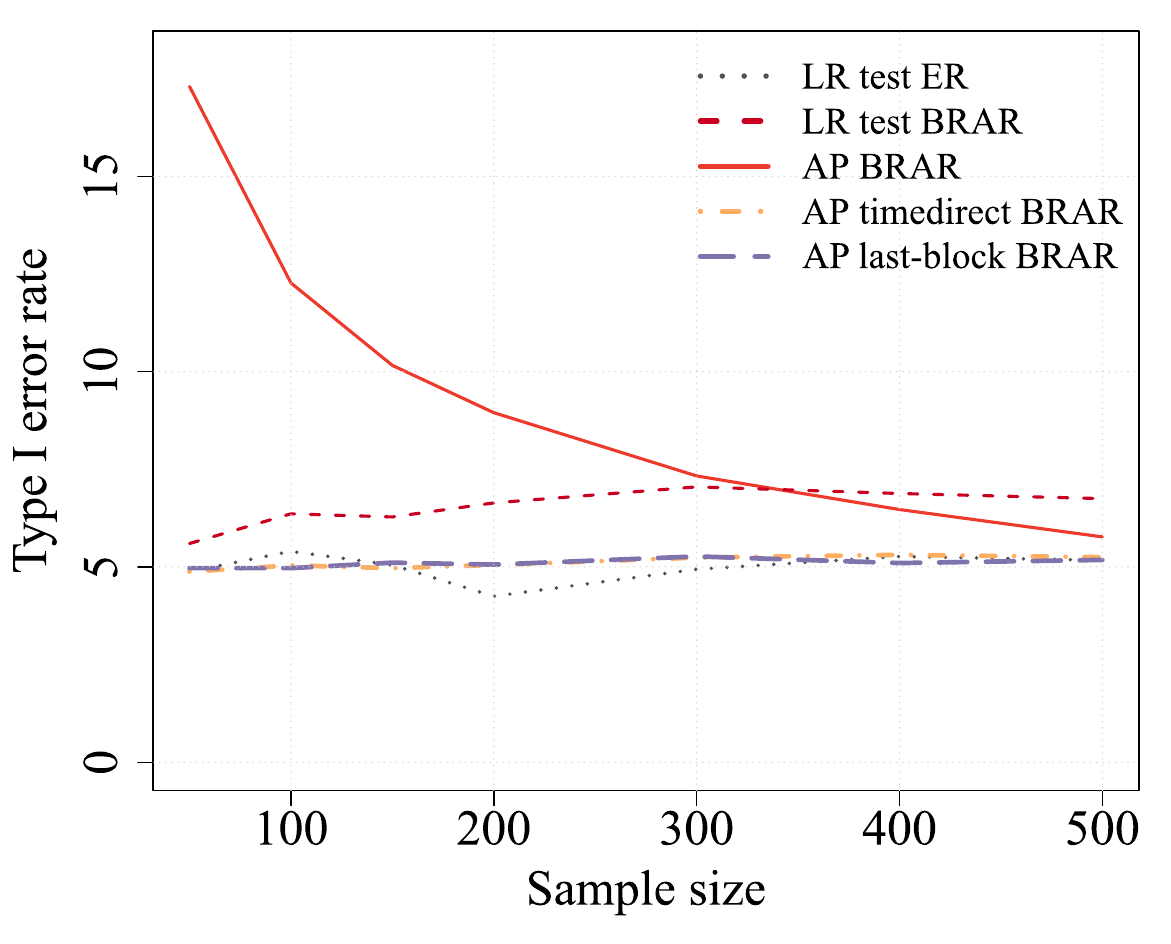}
	\caption{Tuned BRAR.}
    \label{fig:tunedTST1}
\end{subfigure}
\caption{Type I error rate with $B=1$ for (A) standard BRAR and (B) tuned BRAR, Monte Carlo standard errors $\leq 0.005$.}
\label{fig:type1}
\end{figure}

The type I error rates of the different tests as a function of the sample size are presented in Figure~\ref{fig:type1} in two fully sequential designs. Here, we see that the type I error rate for the original AP test varies quite a lot over sample sizes, and that it is quite large. For both the standard and tuned BRAR designs, it does not reach the targeted 5\% even when $N=500$, although the type I error rate is decreasing with sample size. As stated previously, this is due to the discreteness of the test statistic and the aggressiveness of the design. The last-block and timedirect AP test do not have this issue as the test statistics are continuous and can thus take more values. Therefore, the type I error for these versions of the AP test are constant at 5\% for all sample sizes for both designs. The likelihood-ratio test for the ER design also has a targeted 5\% type I error rate for both designs, whereas the likelihood-ratio test for the BRAR designs has an inflated type I error rate, and it is increasing with the sample size for both BRAR designs, although less for the tuned BRAR design. 

\subsubsection{Large-sample behavior}

\begin{figure}[htb!]
\centering
\begin{subfigure}[t]{0.49\linewidth}
    \centering
    \includegraphics[width=\linewidth]{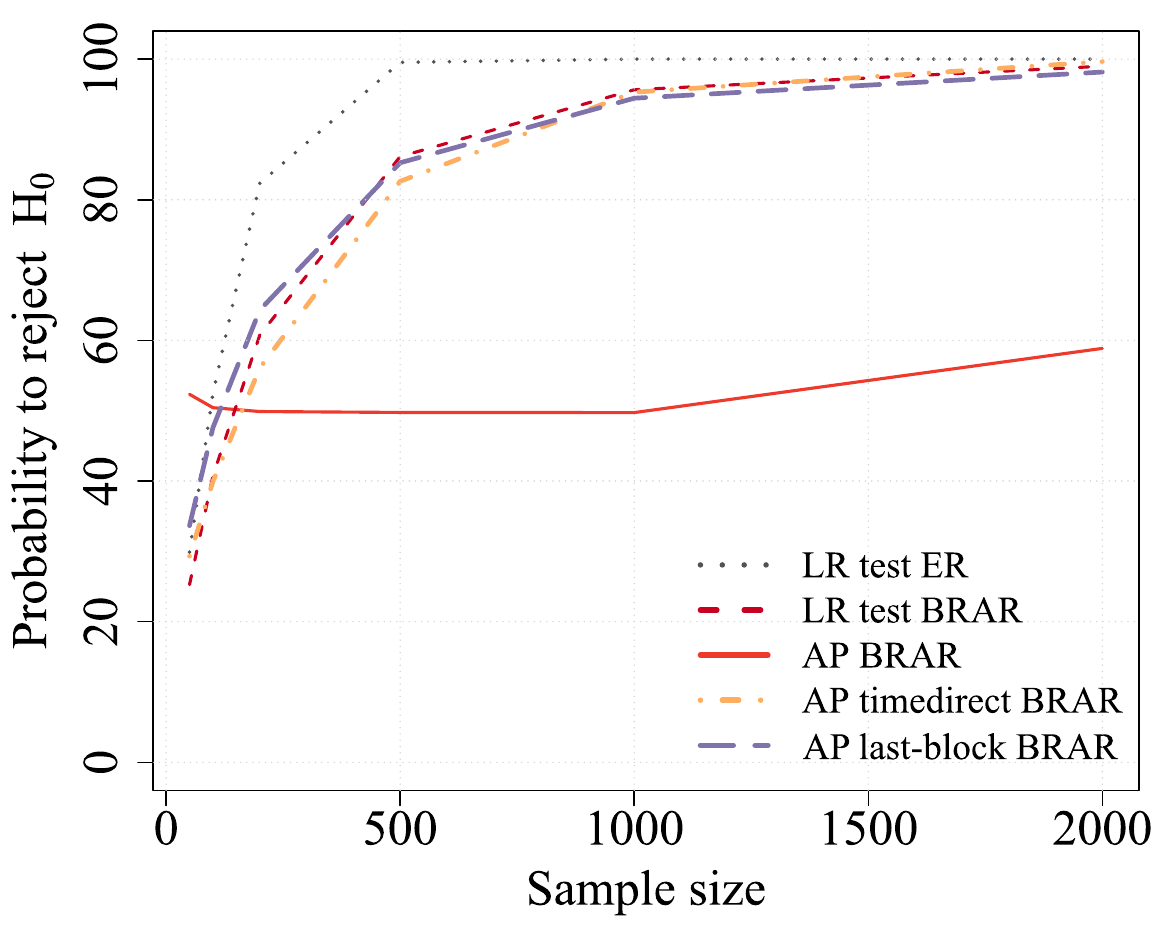}
	\caption{$\lambda_0=1,\;\lambda_1=1.5$.}
    \label{fig:asympModerate}
\end{subfigure}
\begin{subfigure}[t]{0.49\linewidth}
    \centering
    \includegraphics[width=\linewidth]{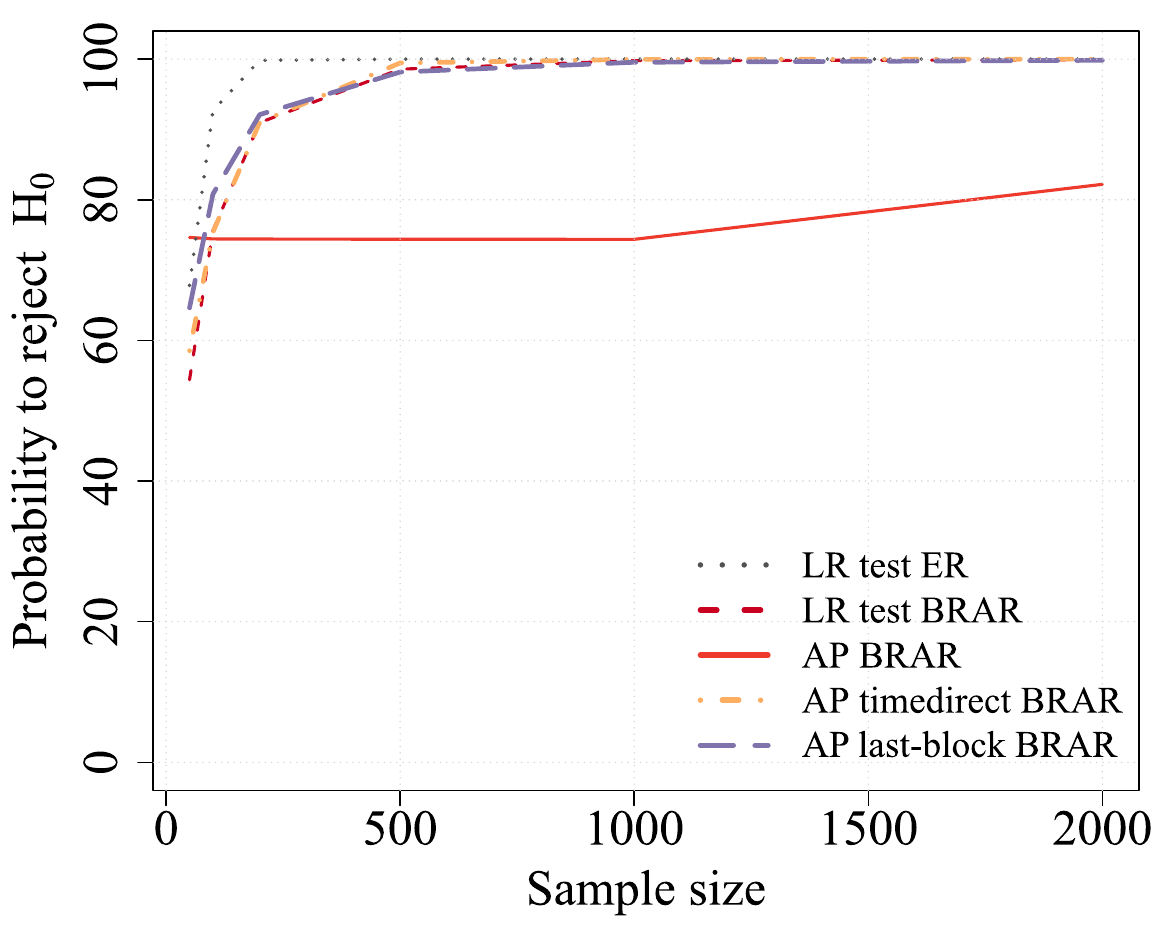}
	\caption{$\lambda_0=1,\;\lambda_1=2$.}
    \label{fig:asympLarge}
\end{subfigure}
\caption{Power curves for (A) $\lambda_0=1,\;\lambda_1=1.5$, and (B) $\lambda_0=1,\;\lambda_1=2$ with $B=1$ as the sample size increase, Monte Carlo standard errors $\leq0.005$.}
\label{fig:asymp}
\end{figure}

The large-sample behavior of the different AP tests are also of interest. Here, we look at how the convergence  of the power for the AP tests as the sample size increases in two different settings: a moderate treatment effect where $\lambda_0=1$ and $\lambda_1=1.5$, and a large treatment effect where $\lambda_0=1$ and $\lambda_1=2$ for the standard BRAR design. The reason that we only look at the standard BRAR design is that the tuned version is very similar to the standard BRAR design as the trial size increase. For a moderate treatment effect, Figure~\ref{fig:asympModerate}, we see that the last-block and timedirect AP tests converge to 100\% at the same rate as the likelihood-ratio test for BRAR design. However, the convergence is slower than the convergence for the likelihood-ratio test for the ER design. The power of the original AP test is almost constant until the sample size 2000 when it is increasing slightly. However, the type I error rate for the original AP test decreases as the sample size increases, Figure~\ref{fig:type1}, and it is only at a sample size of 1000 that the original AP test reaches the targeted 5\%. Thus, when comparing power across the different sample sizes, this should be taken into account. The results for a large treatment effect, Figure~\ref{fig:asympLarge}, are similar. Again, the last-block and timedirect AP tests converge at the same rate as the likelihood-ratio test for the BRAR design, but at a faster rate than for the smaller treatment effect. Again, the power of the original AP test is almost constant until sample size 1000 when it increases, but it is higher than for the smaller treatment effect.

\subsubsection{Other types of outcomes}

In this section, we briefly discuss the performance of the alternative AP tests for other types of outcomes, namely binary and normal with unequal variances. Previous works have investigated the AP test for binary~\citep{barnett2023novel} and normal outcome~\citep{deliu2025efficient}, and here we extend these studies looking at the different alternatives of the AP test. The simulation settings and results are presented in the Supplemental Material. 

Similarly to the studies presented above for exponential endpoints, significant power gains can be made using an alternative AP test functional form for binary and normal outcomes. Again, among the investigated alternatives, the last-block AP test gave the highest power, and controlled the type I error rate, although the timedirect, and other versions, of the AP test provided similar power. For the binary outcome, the last-block and timedirect AP tests for the standard BRAR design have similar power to the Fisher exact test for the ER design, although the last-block and timedirect AP tests have a higher (but still controlled) type I error rate. For the normal outcome, the last-block and timedirect AP tests have lower power than the Z test for the ER design. Furthermore, the Z test for the standard BRAR design has similar power as the last-block AP test, but higher power than the timedirect AP test, with adjustment for strict type I error control. 

As noted in Section \ref{sec:sim}, for exponential outcome in the two arm case, it is only the ratio of the parameter values that affect the BRAR design, and not their absolute values. In the case of normal outcome, the situation depend on whether the variances in the different arms change. Assuming that the variance in each the treatment arm does not change, it is the absolute difference between the treatment means that is relevant, not the values of the treatment means. In other words, a constant shift in the treatment means for the two arms does not change the design. However, if the variances change, the absolute parameter values influence the BRAR design, and in extension also the null and alternative distributions. For binary outcome, we see that the absolute parameter values are always relevant for the design. In these cases, calibration of the critical value under an incorrect null may lead to either inflated or conservative type I error rates (as shown in the Supplemental Material), and care should be taken when calibrating the tests, for instance by analyzing the test in a range of parameter values.

\section{Empirical example} \label{sec:emp}

To illustrate the use and performance of the AP test in an applied context, we consider a simulation study based on \citet{schwartz2004comparison}, a phase 3 study which compares the performance of a new fibrin sealant (Crosseal) with standard topical hemostatic agents among patients after liver resection using standard surgical techniques. The primary outcome was time to hemostasis. There was 121 patients included in the study. The mean time to hemostasis was 282 seconds with Crosseal, compared with 468 seconds with standard agents. Furthermore, it was found that hemostasis was achieved within 10 minutes in 53 patients (91.4\%) treated with the Crosseal and in 44 control patients (69.8\%). Since the time to outcome is short, a RAR design could have been used in this study. We consider two settings, one with the outcome "time to hemostasis" and the other with the binary outcome "hemostasis within 10 minutes". We use the observed estimates as parameter values ($\lambda_0=0.002$ and $\lambda_1=0.0035$, and $p_0=0.7$ and $p_1=0.9$) in the simulations. For both outcome types, we consider two fully sequential designs with 121 patients with 12 people in the burn-in period, one standard BRAR and one tuned BRAR. The priors are a $Gamma(1,0.001)$ for the exponential outcome and $Beta(1,1)$ for the binary outcome. Since this is a phase 3 study, we adjust the critical value of the hypothesis tests to achieve a strict type I error control at 5\%. 

The results for the exponential outcome is presented in Table~\ref{tab:empExExp}. For both designs, we see that the original AP test has a very low power, 26.2\% and 28.4\%, which matches the simulation results in Section~\ref{sec:sim}. For the standard BRAR, the timedirect AP test has a power of 56.4\%, which is higher than both the original AP test and the likelihood-ratio test for the standard BRAR design. As before, the last-block AP test has the highest power, 64.8\%, for the tests in the standard BRAR design. In comparison, the likelihood-ratio test for the ER design has 87.2\% power. On the other hand, the standard BRAR design allocated on average 86\% of patients in the trial to the Crosseal treatment arm as opposed to 50\% with the ER design. This reduces the average time to hemostasis to 315 seconds, in comparison to 376 seconds, i.e. a minute difference. For the tuned BRAR design, all tests show an increase in power compared to the standard BRAR design. The timedirect AP test has a power of 81.2\%, which is still higher than the likelihood-ratio test for the same design which has a power of 75.4\%. Again, the last-block AP test has the highest power of the tests for the tuned BRAR design, 86.6\%. Thus, with the tuned BRAR design, it is possible to get almost equal power as the likelihood-ratio test for the ER design, but with 80\% of patients allocated to the better arm, which reduces the mean time to hemostasis to 330 seconds, i.e. 46 seconds shorter than for the ER design.

The results for the binary outcome "hemostasis within 10 minutes" is presented in Table~\ref{tab:empExBin}. Again, for both designs, we see that the original AP test has low power, 22.0\% and 26.6\%. For the standard BRAR design, the timedirect AP test has an power of 59.3\%, which is higher than the original AP test and similar to the Fisher exact test for the standard BRAR design. As before, the last-block AP test has the highest power, 67.4\%, for the tests in the standard BRAR design. Similarly to the exponential outcome the Fisher exact test for the ER design has the highest power, 88.5\%. However, the standard BRAR design allocated on average 87\% of patients in the trial to the Crosseal treatment arm as opposed to 50\% with the ER design. This increases the number of patients with hemostasis within 10 minutes to 106 compared to 97. 

The results for the tuned BRAR design are similar to those for the exponential outcome. The power for all tests increase for the tuned BRAR design compared to the standard BRAR design, which demonstrates the value of a good RAR design. Again, the last-block AP test has the highest power of the tests for the BRAR design, 82.5\%, compared to the timedirect AP test, 75.8\%, and the Fisher exact test, 79.4\%. None of the tests has the same power as the Fisher exact test for the ER design, 88.5\%, but the tuned BRAR design allocates 79\% of patients to the better arm which results in 104 successes, 7 more people reach hemostasis within 10 minutes on the tuned BRAR design than the ER design. Thus, both power and patient benefit must be taken into account and be weighted against each other when designing a study.

\section{Implementation guidelines and practical recommendations} \label{sec:guide}

To facilitate implementation in practice, we summarise a set of recommendations for applied statisticians and trialists considering the use of the AP test in settings where time trends are not expected to substantially impact the trial results. These guidelines are intended as a pragmatic “how-to” based on our theoretical findings and simulation results. The recommendations are:
    \begin{enumerate}
        \item The original AP test does not offer the highest power gains. Instead, use an alternative functional form. In our simulation studies, we saw an increase in power of between 20\% and 50\% compared to the original version when controlling the type I error, see Figure~\ref{fig:adjpowerPhase3}.
        \item Consider a version of the AP test that weights the allocation probabilities higher in the end as the allocation probabilities for these blocks contain more information, or alternatively, use a higher value of $t_{\mathrm{min}}$.
        \item Include the allocation probabilities for the hypothetical last block in the test statistic. This gives one more data point (containing all data in the trial) in the test, which makes any version of the AP test more powerful.
        \item When using the AP test in the design, investigate the sensitivity of the test statistic to different null hypotheses.
        \item When analyzing an observed dataset, the AP test can be calibrated under the pooled estimate of the parameter. This offers the additional advantage of allowing for an unblinded analysis without necessitating the disclosure of group-specific results, which may be desirable for other operational or regulatory reasons.
        \item In a setting with a time-to-event data with different follow-up times, the time to observation can be used as weights in the AP test statistic as these times contain information as well.
    \end{enumerate}

Choices around the functional form, weighting scheme, and calibration play a central role in determining performance. The principles outlined above provide a structured basis for implementation and can be readily adapted to a wide range of settings. However, the optimal configuration may depend on the specific trial context. The recommendations that we give here are valid when there is no risk for time trends in the data. For trials where temporal shifts are a primary concern, these guidelines should be supplemented with additional trend-robust methods.

\section{Discussion: theoretical implications and performance limits} \label{sec:disc}

In this work, we add further depth to the work on the allocation probability test by \citet{barnett2023novel} and \cite{deliu2025efficient} in a BRAR setting. We introduce the generalized AP test statistic to search for the most powerful definition of the test statistic, and we find that one special case is identical to a Bayesian decision rule. We evaluate the different versions and compare them to other hypothesis tests. These studies aimed to answer three important practical questions stated in the Introduction of the article, and based on the simulation studies, our answers to these questions are:
\begin{enumerate}
    \item How much can the functional form of the AP test increase power for the different outcome measures?
    
    Substantial power, between 20\% and 50\%, gains can be made by changing the functional form of the AP test compared to the original definition. Importantly, the original definition did not give the highest power in any of the investigated settings, and it had varied, and sometimes not bounded by a reasonable target, type I error rates. The last-block AP test, which only uses the allocation probability from the last block, performed best in the settings we investigated for the exponential, binary and normal outcomes, although the timedirect AP test, which weights the allocation probabilities with the block number, performed similarly. Overall, the alternatives that utilized more information about the allocation probabilities and put more emphasis on the probabilities later in the trial performed better and had similar power and type I error rates.
    \item How does the most powerful version of the AP test compare to a
    \begin{enumerate}
        \item non-adaptive design with a traditional analysis?
        
        \hspace*{1em} The performance of the AP test depends heavily on the type of outcome data and whether the adaptive BRAR design is tuned.

        \begin{itemize}
            \item[-] \textit{Exponential Data:} Any version of the AP test in a standard BRAR design is less powerful than the likelihood-ratio test for an ER design. However, if the BRAR design is tuned, the last-block and timedirect AP tests match or exceed the power of the traditional ER design.
            \item[-] \textit{Binary Data (Small Samples):} The last-block and timedirect AP tests in a standard BRAR design can sometimes match or even outperform the power of an ER design analyzed with the Fisher exact test.
            \item[-] \textit{Normal (Continuous, known and unequal variance) Data:} Any version of the AP test in a standard BRAR design is always less powerful than an ER design analyzed with a Z-test.
        \end{itemize}

        \hspace*{1em} In summary, the AP test can compete with or beat traditional non-adaptive designs for binary and exponential data (provided the BRAR design is tuned), but it consistently underperforms traditional designs when dealing with normally distributed data. 
        
        \item BRAR design with a frequentist analysis? 
        
        \hspace*{1em} Similarly to the above, the performance of the AP tests compared to frequentist tests in a BRAR design depend on the design and outcome type.
        
        \begin{itemize}
            \item[-] \textit{Exponential Data:} The last-block and timedirect AP tests always had higher power than the likelihood-ratio test for both the standard and tuned BRAR designs.
            \item[-] \textit{Binary Data (Small Samples):} The last-block and timedirect AP tests match or outperform the power of a Fisher exact test in a standard BRAR design.
            \item[-] \textit{Normal (Continuous, known and unequal variance) Data:} The last-block AP test matches the power of the Z-test for the standard BRAR design.
        \end{itemize}

        \hspace*{1em} In summary, the AP test can compete with or beat traditional frequentist hypothesis tests in a BRAR design for all outcome types. 
        
        \item BRAR design with a Bayesian analysis?
        
        \hspace*{1em} Importantly, in a BRAR design, the Bayesian decision rule can be viewed as a special case of the AP test with extreme weighting (the last-block AP test). The last-block AP test/Bayesian decision rule is based on a sufficient statistic, ensuring that all available information is used. In contrast, the standard frequentist methods had low power because they assume fixed sample sizes and therefore penalize imbalances. Other versions of the AP tests lie between these two approaches: they avoid the frequentist penalty by treating the sequence of treatment allocations as informative evidence. Rather than relying only on final outcomes, some AP tests use the trajectory of allocation probabilities over time, capturing how the algorithm progressively favored one arm as data accumulated. This incorporates information about the adaptive process itself, but because it relies on this trajectory as a proxy rather than the full sufficient statistics, some precision and power are lost compared to the last-block AP test/Bayesian approach.

        \hspace*{1em} Consistent with this interpretation, in all investigated settings the last-block AP test/Bayesian decision rule achieved higher power than the other AP test variants. Moreover, the versions of the AP tests that directly use the allocation probabilities approach the last-block AP test/Bayesian rule as $t_{\mathrm{min}}$ increases, becoming equivalent when $t_{\mathrm{min}} = T+1$. It is also worth highlighting that for the less volatile tuned BRAR design, the performance of the timedirect and last-block AP test/Bayesian decision rule are more similar, which also points to the value of a good design.

    \end{enumerate}
    \item Based on the above, can we make some practical recommendations?

    Beyond the practical guidelines in Section~\ref{sec:guide}, we give some further recommendations and discussions.
    
    First, randomization-based tests are recommended for their control of type I error rates \citep{carter2024regulatory,sverdlov2024sensitivity}, but they greatly reduce power \citep{villar2018response}. Instead, the AP test can be used as a more powerful alternative for a sanity check. Here, we saw that for a BRAR design, the Bayesian decision rule is a special case of the AP test. However, the AP tests also work for any type of RAR design where the allocation probabilities are known, such as the randomized play-the-winner, without the need to specify any prior, in contrast to the Bayesian decision rule. 
    
    Second, the critical values for both the last-block and timedirect AP tests need to be found using simulations, and these can be sensitive to the parameter values. If these are misspecified, the type I error rate can either be inflated or conservative. On the other hand, the original AP test almost always uses the maximum number of blocks that does not result in a zero type I error rate, and is thus less sensitive to simulation assumptions and it uses less computational power, even though it can have a higher type I error rate and lower power than the other two tests. However, any version of the AP test can be calibrated for the observed dataset or be used in re-randomization type, which can decrease the sensitivity to simulation assumptions. 
 
    Ultimately, the analysis method must be tailored to the specific trial design and rigorously evaluated beforehand. Because an optimal method for one design may underperform in another, a one-size-fits-all approach is not recommended. We instead recommend that the hypothesis tests are investigated for the intended study settings. Furthermore, in this work we saw that it is difficult to match the power of an ER design, but it is important to also consider the increase in patient benefit within the trial. In other words, if one wants to keep patient benefit, minimizing power loss is probably better than going for power at the expense of it.
 
\end{enumerate}

In summary, this study investigates alternative forms of the AP test for two BRAR designs, demonstrating that the functional form of the test statistic is important for the statistical power. We also show that the test performance depends on the study design, and therefore, the design and test should be jointly considered. The approach in the paper has been to find construct AP tests for a given design that gives high power. An alternative approach that can be investigated is to find a design that gives a specific test the highest power. Furthermore, the recommendations that we give here are valid in a BRAR setting when there is no concern about time trends in the data. If time trends might be present, different recommendations may be needed. Extending these results to settings with potential time trends is an important area of ongoing work.

\section*{Acknowledgments}

The author gratefully acknowledges Harry Huang for his insightful suggestion regarding the inclusion of a hypothetical final block in the AP test statistic.

\bibliographystyle{plainnat} 
\bibliography{ref}

@article{thompson1933likelihood,
 URL = {http://www.jstor.org/stable/2332286},
 author = {William R. Thompson},
 journal = {Biometrika},
 number = {3/4},
 pages = {285--294},
 title = {On the Likelihood that One Unknown Probability Exceeds Another in View of the Evidence of Two Samples},
 volume = {25},
 year = {1933}
}

@article{barnett2023novel,
    author = {Barnett, Helen Yvette and Villar, Sofía S. and Geys, Helena and Jaki, Thomas},
    title = {A Novel Statistical Test for Treatment Differences in Clinical Trials Using a Response-Adaptive Forward-Looking {G}ittins Index Rule},
    journal = {Biometrics},
    volume = {79},
    number = {1},
    pages = {86-97},
    year = {2023},
    month = {10},
    url = {https://doi.org/10.1111/biom.13581}
}

@article{deliu2025efficient,
    author = {Deliu, Nina and Villar, Sofia S.},
    title = {On the finite-sample and asymptotic error control of a randomization-probability test for response-adaptive clinical trials},
    journal = {Biometrics},
    volume = {81},
    number = {2},
    year = {2025},
    month = {06},
    url = {https://doi.org/10.1093/biomtc/ujaf069},
}

@article{Neyman23,
 author               = {Jerzy Splawa-Neyman},
 comment              = {English translation by D. Dabrowska and T. Speed},
 journal              = {\emph{English translation by D.M. Dabrowska and T.P. Speed in} Statistical Science, \emph{1990}},
 number               = {4},
 pages                = {465--472},
 title                = {On the application of probability theory to agricultural experiments, essay on principles. \emph{Roczniki nauk Rolczych X}, 1-51. {In Polish}},
 volume               = {5},
 year                 = {1923}
}

@article{rubin1974estimating,
  title={Estimating causal effects of treatments in randomized and nonrandomized studies.},
  author={Rubin, Donald B.},
  journal={Journal of Educational Psychology},
  volume={66},
  number={5},
  pages={688-701},
  year={1974},
  publisher={American Psychological Association}
}

@article{rubin1977assignment,
    author = {Donald B. Rubin},
    title ={Assignment to Treatment Group on the Basis of a Covariate},
    journal = {Journal of Educational Statistics},
    volume = {2},
    number = {1},
    pages = {1-26},
    year = {1977}
}

@article{russo2018tutorial,
  title={A tutorial on {T}hompson sampling},
  author={Russo, Daniel J and Van Roy, Benjamin and Kazerouni, Abbas and Osband, Ian and Wen, Zheng and others},
  journal={Foundations and Trends{\textregistered} in Machine Learning},
  volume={11},
  number={1},
  pages={1--96},
  year={2018},
  publisher={Now Publishers, Inc.}
}

@misc{kalkanli2020asymptotic,
      title={Asymptotic Convergence of {T}hompson Sampling}, 
      author={Cem Kalkanli and Ayfer Ozgur},
      year={2020},
      eprint={2011.03917},
      archivePrefix={arXiv},
      primaryClass={cs.LG},
      url={https://arxiv.org/abs/2011.03917}, 
}

@inproceedings{zhang2020advances,
 author = {Zhang, Kelly and Janson, Lucas and Murphy, Susan},
 booktitle = {Advances in Neural Information Processing Systems},
 editor = {H. Larochelle and M. Ranzato and R. Hadsell and M.F. Balcan and H. Lin},
 pages = {9818--9829},
 publisher = {Curran Associates, Inc.},
 title = {Inference for Batched Bandits},
 url = {https://proceedings.neurips.cc/paper\_files/paper/2020/file/6fd86e0ad726b778e37cf270fa0247d7-Paper.pdf},
 volume = {33},
 year = {2020}
}

@article{villar2018response,
  title={Response-adaptive designs for binary responses: how to offer patient benefit while being robust to time trends?},
  author={Villar, Sof{\'\i}a S. and Bowden, Jack and Wason, James},
  journal={Pharmaceutical statistics},
  volume={17},
  number={2},
  pages={182--197},
  year={2018},
  publisher={Wiley Online Library}
}

@misc{kaddaj2025thompson,
      title={Thompson, {U}lam, or {G}auss? {M}ulti-criteria recommendations for posterior probability computation methods in Bayesian response-adaptive trials}, 
      author={Daniel Kaddaj and Lukas Pin and Stef Baas and Edwin Y. N. Tang and David S. Robertson and Sofía S. Villar},
      year={2025},
      eprint={2411.19871},
      archivePrefix={arXiv},
      primaryClass={stat.ME},
      url={https://arxiv.org/abs/2411.19871}, 
}

@misc{pin2025github,
      title={Clinical Trials Using Response Adaptive Randomization}, 
      author={Lukas Pin and Maja Neubauer and David S. Robertson and Stef Baas and Sofía S. Villar},
      year={2025},
      version={1.0.0},
      url={https://github.com/lukaspinpin/RA-ClinicalTrials}, 
      urldate={2025-07-25}
}

@article{rubenstein2005design,
author = {Rubinstein, Lawrence V. and Korn, Edward L. and Freidlin, Boris and Hunsberger, Sally and Ivy, S. Percy and Smith, Malcolm A. },
title = {Design Issues of Randomized Phase {II} Trials and a Proposal for Phase {II} Screening Trials  },
journal = {Journal of Clinical Oncology},
volume = {23},
number = {28},
pages = {7199-7206},
year = {2005},
URL = {https://ascopubs.org/doi/abs/10.1200/JCO.2005.01.149}
}

@article{schwartz2004comparison,
  author = {Schwartz, Myron and Madariaga, Juan and Hirose, Ryutiao and Shaver, Timothy R. and Sher, Linda and Chari, Ravi and Colonna, John O., II and Heaton, Nigel and Mirza, Darius and Adams, Reid and Rees, Myrddin and Lloyd, David},
  title = {Comparison of a New Fibrin Sealant With Standard Topical Hemostatic Agents},
  journal = {Archives of Surgery},
  volume = {139},
  number = {11},
  pages = {1148-1154},
  year = {2004},
  month = {11},
  url = {https://doi.org/10.1001/archsurg.139.11.1148}
}

@article{morris2019using,
  author = {Timothy P. Morris and Ian K. White and Michael J. Crowther},
  title = {Using simulation studies to evaluate statistical methods},
  journal = {Statistics in Medicine},
  year = {2019},
  volume = {38},
  number = {11},
  pages = {2074--2102},
  url = {https://doi.org/10.1002/sim.8086}
}

@article{carter2024regulatory,
 author = {Kerstine Carter and Annika L. Scheffold and Jone Renteria and Vance W. Berger and Yuqun Abigail Luo and Jonathan J. Chipman and Oleksandr Sverdlov},
 title = {Regulatory Guidance on Randomization and the Use of Randomization Tests in Clinical Trials: A Systematic Review},
 journal = {Statistics in Biopharmaceutical Research},
 volume = {16},
 number = {4},
 pages = {428--440},
 year = {2024},
 URL = {https://doi.org/10.1080/19466315.2023.2239521}
}

@article{sverdlov2024sensitivity,
  title={On "{R}e-randomization tests as sensitivity analyses to confirm immunological noninferiority of an investigational vaccine: Case study" by {L}uca {G}rassano et al.(2023), {P}harmaceutical {S}tatistics.},
  author={Sverdlov, Oleksandr and Berger, Vance W and Carter, Kerstine},
  journal={Pharmaceutical Statistics},
  volume={23},
  number={3},
  year={2024}
}

\begin{table}[htb!]
    \centering
    \caption{Percentage of patients on the correct treatment, standard deviation in parenthesis, and average number of total successes for BRAR compared with ER at $N=100$ and $N=500$. True treatment ratio of 50\%.}
    \label{tab:patben}
    \begin{tabular}{l S[table-format=2.2] S[table-format=2.2] S[table-format=2.2]}
        \toprule
        & {\textbf{Standard BRAR}} & {\textbf{Tuned BRAR}} & {\textbf{ER}} \\
        \midrule
        \multicolumn{4}{l}{\textit{\% Patients on correct treatment}} \\
        \midrule
        $\boldsymbol{N=100}$ & 79\!\!\!\!(14.0) & 73\!\!\!\!(11.2) & 50 \\
        $\boldsymbol{N=500}$ & 91\!\!\!\!(4.3) & 86\!\!\!\!(4.7) & 50 \\
        \midrule
        \multicolumn{4}{l}{\textit{Average time to event}} \\
        \midrule
        $\boldsymbol{N=100}$ & 0.74 & 0.76 & 0.81 \\
        $\boldsymbol{N=500}$ & 0.70 & 0.72 & 0.81 \\
        \bottomrule
    \end{tabular}
\end{table}

\begin{table}[htb!]
    \centering
    \caption{Power and patient benefit for the different hypothesis tests for the empirical example with exponential outcome, standard deviation of patient benefit given in parenthesis. Monte Carlo standard errors $\leq 0.005$.}
    \begin{tabular}{lccc}
        \toprule
        & \begin{tabular}[c]{@{}l@{}}Power\end{tabular} 
        & \begin{tabular}[c]{@{}l@{}}Patients on \\ correct treatment\end{tabular}
        & \begin{tabular}[c]{@{}l@{}}Average time \\ to event\end{tabular} \\ 
        \midrule

        \multicolumn{4}{l}{\textit{Standard BRAR}} \\
        \midrule
        $\boldsymbol{\mathrm{AP}_T}$ & 26.2\% & 86\% (9.0\%) & 315  \\ \hdashline
        $\boldsymbol{\mathrm{AP}_T^{timedirect}}$ & 66.4\% & 86\% (9.0\%) & 315 \\ \hdashline
        \textbf{LR BRAR} & 57.7\% & 86\% (9.0\%) & 315 \\ \hdashline
        $\boldsymbol{\mathrm{AP}_T^{lastblock}}$ & 73.2\% & 86\% (9.0\%) & 315 \\ \hdashline 
        \textbf{LR ER} & 87.2\% & 50\% & 376 \\ 

        \midrule
        \multicolumn{4}{l}{\textit{Tuned BRAR}} \\
        \midrule
        $\boldsymbol{\mathrm{AP}_T}$ & 28.4\% & 80\% (7.8\%) & 330  \\ \hdashline
        $\boldsymbol{\mathrm{AP}_T^{timedirect}}$ & 81.2\% & 80\% (7.8\%) & 330 \\ \hdashline
        \textbf{LR BRAR} & 75.4\% & 80\% (7.8\%) & 330 \\ \hdashline
        $\boldsymbol{\mathrm{AP}_T^{lastblock}}$ & 86.6\% & 80\% (7.8\%) & 330 \\ \hdashline 
        \textbf{LR ER} & 87.2\% & 50\% & 376 \\ 

        \bottomrule
    \end{tabular}
    \label{tab:empExExp}
\end{table}

\begin{table}[htb!]
    \centering
    \caption{Power and patient benefit for the different hypothesis tests for the empirical example with binary outcome.}
    \begin{tabular}{lccc}
        \toprule
        & \begin{tabular}[c]{@{}l@{}}Power\end{tabular}
        & \begin{tabular}[c]{@{}l@{}}Patients on \\ correct treatment\end{tabular}
        & \begin{tabular}[c]{@{}l@{}}Average number \\ of successes\end{tabular} \\
        \midrule

        \multicolumn{4}{c}{\textit{Standard BRAR}} \\
        \midrule
        $\boldsymbol{\mathrm{AP}_T}$ & 22.0\% & 87\% (10.2\%) & 106 \\ \hdashline
        $\boldsymbol{\mathrm{AP}_T^{timedirect}}$ & 59.3\% & 87\% (10.2\%) & 106 \\ \hdashline
        \textbf{Fisher BRAR} & 60.6\% & 87\% (10.2\%) & 106 \\ \hdashline
        $\boldsymbol{\mathrm{AP}_T^{lastblock}}$ & 67.4\% & 87\% (10.2\%) & 106 \\ \hdashline
        \textbf{Fisher ER} & 88.5\% & 50\% & 97 \\

        \midrule
        \multicolumn{4}{c}{\textit{Tuned BRAR}} \\
        \midrule
        $\boldsymbol{\mathrm{AP}_T}$ & 26.6\% & 79\% (8.2\%) & 104 \\ \hdashline
        $\boldsymbol{\mathrm{AP}_T^{timedirect}}$ & 75.8\% & 79\% (8.2\%) & 104 \\ \hdashline
        \textbf{Fisher BRAR} & 79.4\% & 79\% (8.2\%) & 104 \\ \hdashline
        $\boldsymbol{\mathrm{AP}_T^{lastblock}}$ & 82.5\% & 79\% (8.2\%) & 104 \\ \hdashline
        \textbf{Fisher ER} & 88.5\% & 50\% & 97 \\

        \bottomrule
    \end{tabular}
    \label{tab:empExBin}
\end{table}

\end{document}